\documentclass[preprint, amsmath,amssymb, superscriptaddress, aps,prb,floatfix]{revtex4-1}

\usepackage{graphicx}
\usepackage[T1]{fontenc}
\usepackage{lmodern}
\usepackage{color}
\usepackage{natbib}
\usepackage{amsmath}

\begin{document}


\title{Quantum oscillations in an optically-illuminated two-dimensional electron system at the LaAlO$_3$/SrTiO$_3$ interface}

\author{I. Leermakers}
\altaffiliation [Present address: ]{Applied physics, TNW, Fontys University of Applied Sciences, Eindhoven, The Netherlands}
\affiliation{High Field Magnet Laboratory (HFML-EMFL) and Institute for Molecules and Materials, Radboud University, Nijmegen, The Netherlands}

\author{K. Rubi}
\email{Rubi.Km@ru.nl}
\affiliation{High Field Magnet Laboratory (HFML-EMFL) and Institute for Molecules and Materials, Radboud University, Nijmegen, The Netherlands}

\author{M. Yang}
\altaffiliation [Present address: ]{Wuhan National Magnetic Field Center, Huazhong University of Science and Technology, Wuhan 430074, P. R. China}
\affiliation{Laboratoire National des Champs Magn\'{e}tiques Intenses (LNCMI-EMFL), CNRS, Univ. Grenoble Alpes, INSA-T, UPS,  Toulouse, France}

\author{B. Kerdi}
\affiliation{Laboratoire National des Champs Magn\'{e}tiques Intenses (LNCMI-EMFL), CNRS, Univ. Grenoble Alpes, INSA-T, UPS,  Toulouse, France}

\author{M. Goiran}
\affiliation{Laboratoire National des Champs Magn\'{e}tiques Intenses (LNCMI-EMFL), CNRS, Univ. Grenoble Alpes, INSA-T, UPS,  Toulouse, France}

\author{W. Escoffier}
\affiliation{Laboratoire National des Champs Magn\'{e}tiques Intenses (LNCMI-EMFL), CNRS, Univ. Grenoble Alpes, INSA-T, UPS,  Toulouse, France}

\author{A. S. Rana}
\altaffiliation [Present address: ]{School of Engineering and Technology, BML Munjal University (Hero Group), Gurgaon, India – 122413}
\affiliation{MESA+ Institute for Nanotechnology, University of Twente, P.O. Box 217, 7500 AE Enschede, The Netherlands}

\author{A. E. M. Smink}
\altaffiliation [Present address: ]{Max Planck Institute for Solid State Research, Heisenbergstraße 1, 70569 Stuttgart, Germany}
\affiliation{MESA+ Institute for Nanotechnology, University of Twente, P.O. Box 217, 7500 AE Enschede, The Netherlands}

\author{A. Brinkman}
\author{H. Hilgenkamp}
\affiliation{MESA+ Institute for Nanotechnology, University of Twente, P.O. Box 217, 7500 AE Enschede, The Netherlands}

\author{J. C. Maan}
\affiliation{High Field Magnet Laboratory (HFML-EMFL) and Institute for Molecules and Materials, Radboud University, Nijmegen, The Netherlands}
\author{U. Zeitler}
\email{Uli.Zeiter@ru.nl}
\affiliation{High Field Magnet Laboratory (HFML-EMFL) and Institute for Molecules and Materials, Radboud University, Nijmegen, The Netherlands}

\date{\today}

\begin{abstract}

We have investigated the illumination effect on the magnetotransport properties of a two-dimensional electron system at the LaAlO$_3$/SrTiO$_3$ interface. The illumination significantly reduces the zero-field sheet resistance, eliminates the Kondo effect at low-temperature, and switches the negative magnetoresistance into the positive one. A large increase in the density of high-mobility carriers after illumination leads to quantum oscillations in the magnetoresistance originating from the Landau quantization. The carrier density ($\sim 2 \times 10^{12}$ cm$^{-2}$) and effective mass ($\sim 1.7 ~m_e$) estimated from the oscillations suggest that the high-mobility electrons occupy the d$_{xz/yz}$ subbands of Ti:t$_{2g}$  orbital extending deep within the conducting sheet of SrTiO$_3$. Our results demonstrate that the illumination which induces additional carriers at the interface can pave the way to control the Kondo-like scattering and study the quantum transport in the complex oxide heterostructures.

\end{abstract}

\maketitle


\section{Introduction}

The discovery of a conducting interface between the two band insulators LaAlO$_3$ (LAO) and SrTiO$_3$ (STO) has opened a new research field of oxide electronics.\cite{Ohtomo2004} 
Considerable efforts have been made to understand the mechanisms for the formation of this two-dimensional electron system (2DES) and to gain control over its electronic properties. 
Specifically, depending on the growth parameters, properties such as high mobility, \cite{McCollam2014, Xie2014} superconductivity,\cite{Caviglia2008, Gariglio2009, Richter2013, Prawiroatmodjo2016} magnetism \cite{Brinkman2007, Bert2011, Gorkov2015} or a combination of these, can be observed. 
These phenomena are interesting for fundamental research, as well as for novel oxide electronics applications.\cite{Bal2015, Goswami2016, Liang2013}

One of the major challenges, however, is to predict the exact properties after a specific growth procedure,\cite{Brinkman2007, Xu2016} and to manipulate them.
The most common tool in the latter category is back- or top-gating experiments. 
The gating affects the Fermi energy and the confinement potential, thereby changing the electron concentration, shifting the critical temperature of the superconductivity and/or changing the magnetic properties in LAO/STO.\cite{Caviglia2008, Caviglia2010, Fete2012}
Another way to manipulate the electronic state of the material is by illumination. 
It has been shown that UV-light influences the 2DES significantly, for example, it can lower the resistivity by orders of magnitude due to the large enhancement in carrier density ($n$) and mobility ($\mu$).\cite{Guduru2013, DiGennaro2013, Lu2014, Liu2016b} Despite optically induced high-mobility carriers, quantum oscillations in the illuminated LAO/STO samples were not observed previously in magnetotransport experiments,\cite{Guduru2013} most likely because of insufficient magnetic field and temperature required for Landau quantization.


In this paper, we have studied the magnetotransport properties of the illuminated LAO/STO heterostructure, with 26 unit cells of LAO grown on a TiO$_2$-terminated STO (100) substrate, in high magnetic fields and low temperatures. This particular heterostructure grown with slightly different growth parameters has been systematically investigated and primary features such as Kondo effect,\cite{Brinkman2007} anisotropic magnetoresistance, \cite{wang2011} and multiband conduction \cite{Guduru2013b, Guduru2013, Jost2015} have been reported. Here, we observed that the illumination completely eradicates the Kondo-like features at low temperatures ($< 15$ K) and switches the magnetoresistance from negative ($-20 \%$) to positive ($+60\%$). Interestingly, very small amplitude Shubnikov-de Haas (SdH) oscillations, barely visible in the raw data, superimpose the positive magnetoresistance. Executing two-band Drude model analysis to the non-linear Hall resistance, we reveal the coexistence of low- and high-mobility electron channels at the interface.  More specifically, the high-mobility carriers exhibit a small density ($\sim 2~\times~10^{12}$ cm$^{-2}$) which is comparable to that estimated from the SdH oscillations' frequency. Furthermore, the cyclotron mass ($\sim 1.7~m_e$) estimated from the oscillations' amplitude suggests that the high-mobility electrons occupy heavy subbands d$_{xz/yz}$ of Ti:3d band existing away from the interface and extending deep within the conducting sheet.

\section{Experimental methods and results}
\subsection{Sample characterization}
The epitaxial LAO film of 10 nm (26 u.c.) was grown on a TiO$_2$-terminated STO substrate using pulsed laser deposition at temperature of $850$~$^\circ$C, oxygen pressure of $1.5\times10^{-2}$~mbar, and laser fluency of $1.3$~J/cm$^2$. The growth of LAO was monitored by means of \emph{in-situ} reflective high-energy electron diffraction (RHEED). As shown in Fig.1, the observed RHEED oscillations specified a layer-by-layer growth of 26 u.c. of LAO on the STO substrate.

\begin{figure}
\centering
\includegraphics[width=\linewidth]{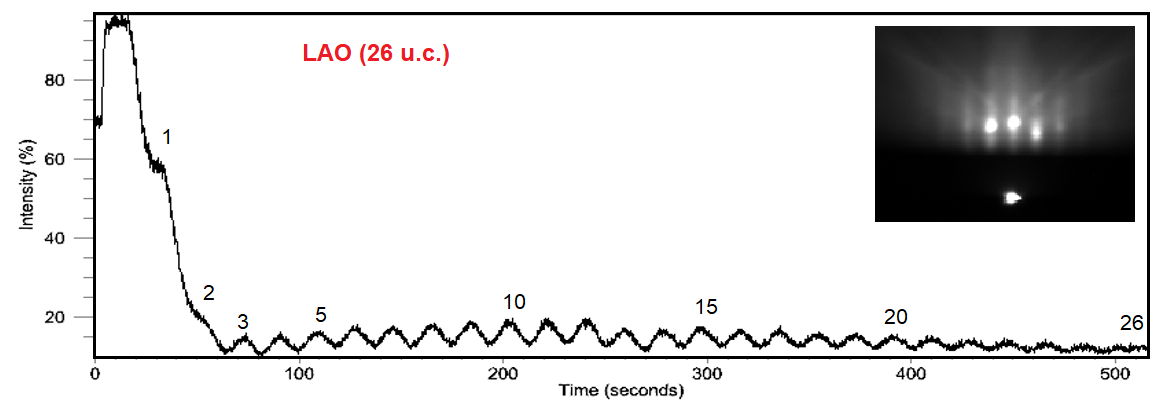}
\caption {RHEED oscillations for the LAO layers grown on a STO substarte. The inset shows the RHEED pattern after growth of 26 u.c. LAO layers. The maxima in the RHEED intensity clearly show the controlled layer-by-layer growth with the number of layers indicated on top of them.}
\end{figure}

\begin{figure}
\centering
\includegraphics[width=14 cm]{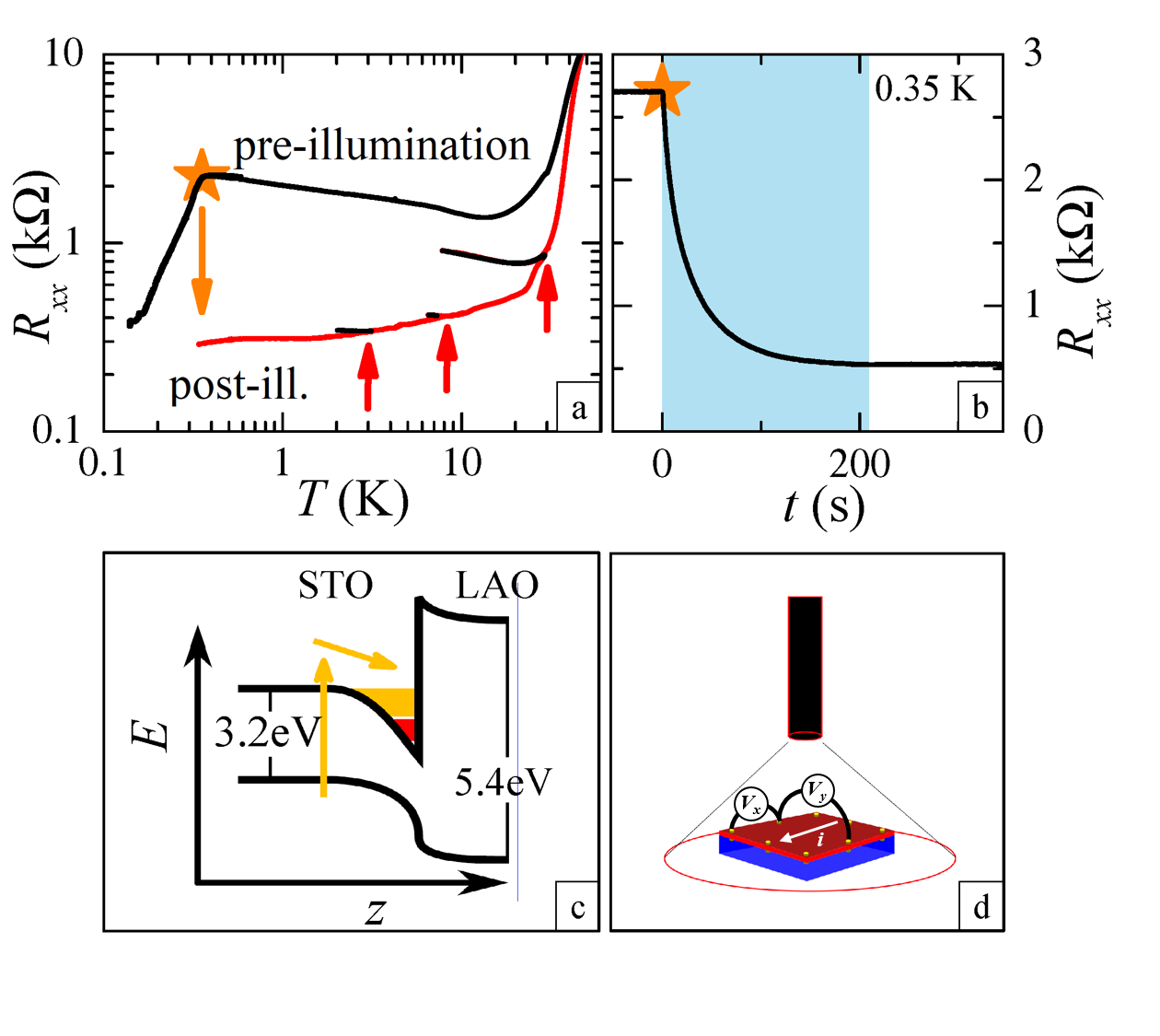}
\caption {(a) Resistance $R_{xx}$ versus temperature $T$ on a log-log scale. 
The high resistance  cool-down curve is represented by a black line.   
At the orange star, the sample is illuminated, visible in the time dependent measurement in (b). 
The post-illumination low-resistance warm-up curve is shown in red.
At three temperatures ($3$, $8$, $30$ K) the warm-up (red) is interrupted by a partial cool-down (black). 
(b) $R$ as a function of  time $t$ while illuminating with an intensity of few $\mu$W (blue area). 
(c) A sketch of the band bending at the LAO/STO interface. Arrows indicate excited electrons.
(d) A schematic diagram of the illumination setup (not to scale) with the contacts used for resistance measurements.}
\label{fig:illR}
\end{figure}

For magnetotransport measurements, we use both dc-field (up to $30$~T) at the High Field Magnet Laboratory in Nijmegen and pulsed-field (up to $55$~T) at the Laboratoire National des Champs Magn\'{e}tiques Intenses in Toulouse. The sample was mounted on a ceramic chip-carrier and electrically contacted in van der Pauw geometry with aluminum wires using an ultrasonic wire-bonder. We used a custom-designed probe suitable for illumination at low temperatures. While the illumination and magnetotransport measurements were performed at $T$ = 0.35 K in a $^3$He system, the temperature dependence of zero-field resistance of the pristine sample was also measured in a dilution fridge down to 0.15 K. The \emph{in-situ} illumination in the dilution fridge was not possible in the setup used because of no optical access.  
The excitation current of  $1$~$\mu$A was passed through two contacts on the middle of the sample (see Fig.~\ref{fig:illR}d) and voltages were measured on both sides parallel to the current path for the longitudinal resistance $R_{xx}$, and perpendicular to the current path for the Hall resistance $R_{yx}$. Due to some inhomogeneity, there were slight differences between the $R_{xx}$ signal, measured on the different sides of the sample. In order to avoid geometric admixtures, we measured $R_{xx}$ and $R_{yx}$ for both positive and negative magnetic field directions and calculated the symmetrised $R_{xx}$ and antisymmetrised $R_{yx}$ using the formula  $R_{xx}= \frac{R_{xx}(+B) + R_{xx}(-B)}{2}$ and $R_{yx} = \frac{R_{yx}(+B)-R_{yx}(-B)}{2}$, respectively.

Mounting the sample on our measurement setup requires it to be in the open air and light for several minutes. This means that the sample gets enough UV-light at room temperature to possibly be in an excited state before the experiment starts. To minimise this effect, after mounting, the sample was kept in the dark at room temperature for several hours, allowing the system to recover to its ground state before start cooling. Fig.~\ref{fig:illR}(a) shows the temperature dependence of $R_{xx}$ before and after illuminating the sample. Before illumination, $R_{xx}$ (black curve) was measured while cooling the sample from room temperature to $0.35$~K.  From the black curve, $R_{xx}$ decreases from a several~M$\Omega$ through a minimum around $15$~K, after which it starts to increase logarithmically with decreasing $T$, reaching 2.7~k$\Omega$ around $0.35$~K. This logarithmic behavior is typically assigned to Kondo scattering on the Ti$^{3+}$ ions \cite{Brinkman2007, Guduru2013, Jost2015} or magnetic moments from oxygen vacancies.\cite{Li2015c} 
Measuring the same sample at lower temperatures (< $0.35$~K) in the dilution fridge, we noticed an additional rapid drop in $R_{xx}$, most likely related to the onset of a superconducting phase at low $T$ ($\sim$ 0.10 K) \cite{Reyren2007a}. However, no full superconducting phase transition is detected in the temperature range down to 0.15 K.

\subsection{Illumination}

The illumination was performed at $T$ = 0.35 K in $^3$He system with a $3.3$~eV ($373$~nm) laser, which is slightly higher than the bandgap energy of STO (3.2 eV). 
The sketch in Fig.~\ref{fig:illR}(d) shows how the light is brought to the sample by a multimode optical fiber. 
This fiber terminates $2$~cm above the sample which is far enough away for the light to diverge and illuminate the entire surface of the sample homogeneously. 
When the light reaches the sample, it is first transmitted through the LAO, which has a bandgap energy of $5.4$~eV;
and then reaches the STO where it excites electrons. 
Our understanding is that the electrons are transported to, and trapped at the interface due to the confinement potential.
The electrons follow the arrows drawn in the band-bending sketch of Fig.~\ref{fig:illR}(c) (not to scale), while the simultaneously created holes move in the opposite direction.

In Fig.~\ref{fig:illR}(b), the effect of the illumination on the sample resistance $R_{xx}$ is shown as a function of time. 
The start temperature and time of the illumination are indicated with a star in Fig.~\ref{fig:illR}(a) and (b), respectively.
After a stable value prior to the illumination (here $2.7$~k$\Omega$, but in general the precise value depends on the cool-down history and the light-exposure history of the sample), the resistance drops within $200$~seconds by nearly one order of magnitude, depending on the incident power (blue shaded area).
Once the illumination is stopped, the resistance remains stable for several hours as long as the sample is kept at low temperature. If the illumination power is increased, the resistance reaches a slightly lower value, however, as soon as the light is turned off, the resistance increases and stabilizes to a persistent value that is only weakly dependent on the incident illumination power. For instance, in  Fig.~\ref{fig:illR}(b), we observe a 1\% increase of resistance after the light is turned off which is barely visible on the scale of the figure.
The remaining conductivity is then persistent for more than several hours. The effects of subsequent heating on the resistance is shown in Fig.~\ref{fig:illR}(a) (red curve), which is of different character to that obtained pre-illumination (black curve). 
The post-illumination sample shows metallic behavior with a monotonically rising $R_{xx}(T)$ wihtout any Kondo-like features.  
At three temperatures, $3$~K, $8$~K and $30$~K, the warm-up was intentionally interrupted by a partial cool-down (black lines)  to verify stability.
As seen in the figure, the resistance already recovers partially at each of the partial cool-downs, and substantially at $30$~K.
The fact that the system recovers to its original state when warming above $50$~K for a substantial time period (many hours), suggests that the persistent conductivity is related to the 2DES observed at low temperature.

The decrease of resistivity during illumination suggests that the subsequent persistent conductivity is due to either an increase of $n\cdot\mu$ in the existing channels, or, alternatively, the creation of an additional channel with significantly higher $n\cdot\mu$. While the first scenario can be exemplified by analyzing magnetotransport data in terms of a two-channel Drude model, the second scenario requires a dedicated theoretical calculation which is beyond the scope of this work. In next section, we discuss the magnetic field dependence of $R_{xx}$ and $R_{yx}$ for pre- and post-illumination and analyze the data through Drude model for two parallel channels. 

\subsection{Magnetotransport}
\begin{figure}
\centering
\includegraphics[width=14 cm]{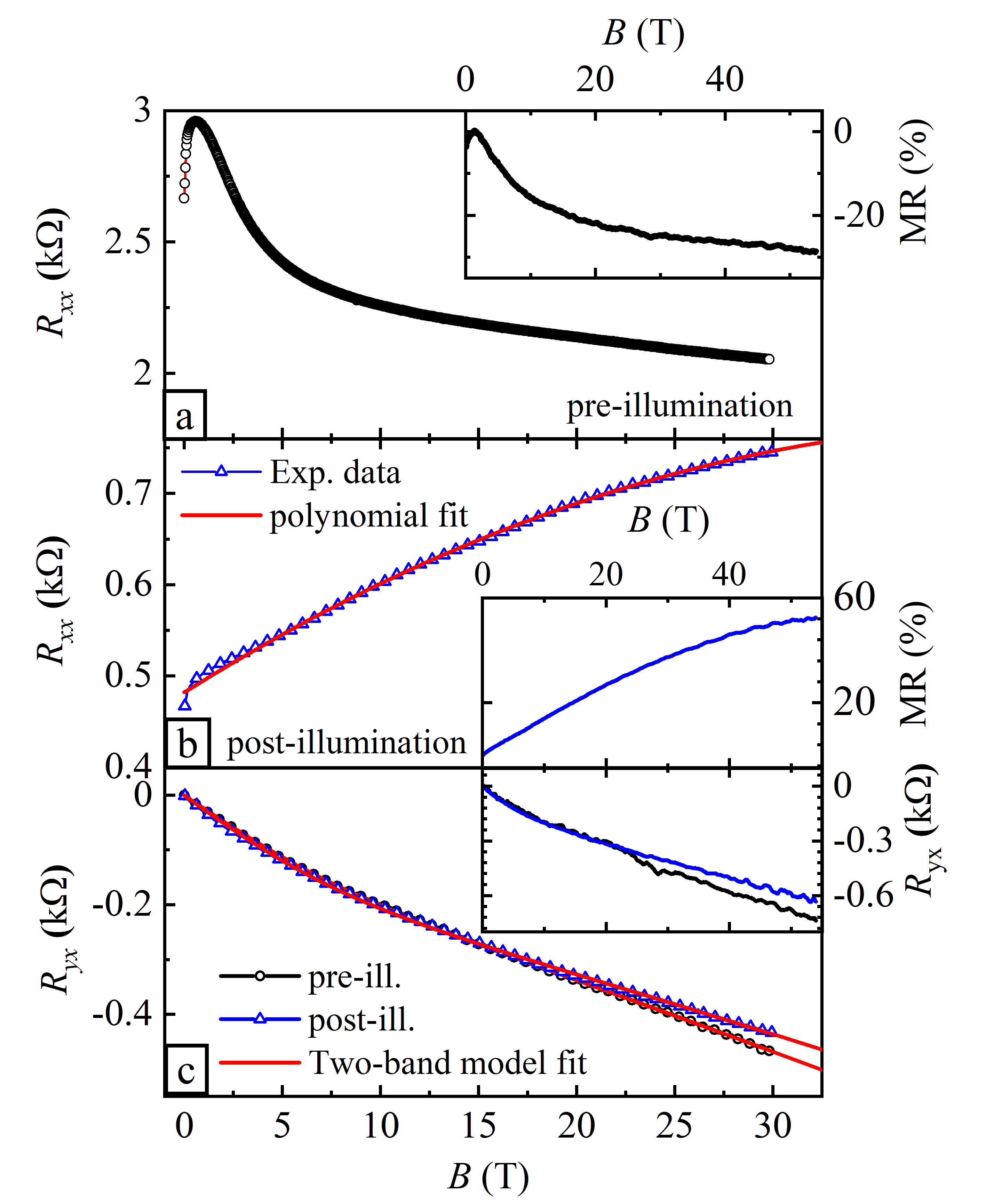}
\caption {Magnetic field dependence of $R_{xx}$ from (a) pre-illumination and (b) post-illumination measurements at $T = 0.35$~K and upto $B$ = 30 T. (c) Hall resistances $R_{yx}(B)$ from pre-illumination (black circle) and post-illumination (blue triangle). While the solid line (red color) in (b) is a second order polynomial fit, the lines in (c) are two-band Drude model fits.  Insets show the magnetoresistance {\sl MR}~$ = \frac{R(B)-R(0)}{R(0)}$ and the Hall resistance $R_{yx}$ measured pre- and post-illumination in pulsed magnetic fields up to $55$~T and at $T$ = 0.50 K.}
\label{fig:MR}
\end{figure}

The pre-illumination value of $R_{xx}$ measured while sweeping the dc-field from 0 to 30 T at $T=0.35$~K is shown in the main panel of Fig.~\ref{fig:MR}(a). 
A rapid increase in low fields, below $1~T$, is most probably due to low-resistance phase (onset of superconducting phase) which is destroyed by a small magnetic field.
For $B>1$~T, a pronounced negative magnetoresistance is observed, which can be attributed to a reduction of spin scattering on magnetically oriented Ti$^{3+}$ ions.\cite{Brinkman2007}
From post-illumination (Fig.~\ref{fig:MR}(b)), $R_{xx}$ drops by almost an order of magnitude and the negative MR is no longer observable, instead, a positive MR now dominates. Interestingly, we also observed very small amplitude oscillations superimposed on the positive MR for post-illumination, which will be discussed in detail later. The insets in Fig.~\ref{fig:MR} show data measured in pulsed magnetic field up to 55 T at $T$ = 0.5 K. The data obtained in pulsed-field are in line with those measured in dc-field, thereby evidencing the reproducibility of the main sample's charateristics. From the inset of Fig.~\ref{fig:MR}(b), the positive MR starts to saturate at high fields $\sim 50$~T (inset in Fig.~\ref{fig:MR}(b)), which is typical behavior for a multi-channels system with channels of different carrier density \cite{sondheimer1947theory}.  To extract the charge carrier's properties from the Hall-effect and the quantum oscillations, we focus on the dc-field data, which has a higher signal-to-noise ratio.

We compare the Hall resistance $R_{yx}$ from pre- and post-illumination measurements in Fig.~\ref{fig:MR}(c). Unlike to the previously reported linear $R_{yx}(B)$ before illumination,\cite{Guduru2013, Guduru2013b} we observed a nonlinear $R_{yx}(B)$ in low-field regime (0-15 T) and a linear $R_{yx}(B)$ in high-field regime ($B~>$ 15 T). The distinct $R_{yx}(B)$ behavior for the same heterostructure, 26 uc-LAO/STO, could be because of slightly different growth parameters, particularly, the oxygen pressure and the growth temperature which significantly affect the carrier density and mobility in the 2DES \cite{wang2011, Hernandez2012, han2016}. Interestingly, the illumination does merely affect the Hall resistance at low magnetic fields which only starts to deviate from its pre-illumination behavior in fields above $\sim 20$~T. This simple observation underlines the importance of high magnetic fields for a proper two conduction-channel analysis of the Hall resistance of this sample.
Keeping in mind that the Drude-model may not give full and reliable qualitative information of multichannels, we can still use it to quantify the properties of two channels of different carrier densities, $n_{1,2}$ and mobilities $\mu_{1,2}$. According to two-band Drude model, the zero-field sheet resistance $R_s(0)$ and the magnetic field dependence of $R_{yx}$ and $R_{xx}$ are defined as

\begin{equation}
R_s (0) = \frac{1}{e} \left(\frac{1}{n_1\mu_1+n_2\mu_2}\right),
\end{equation}

\begin{equation}
R_{yx}(B)=-\frac{B}{e}\frac{(n_1\mu_1^2+n_2\mu_2^2)+(\mu_1\mu_2B)^2(n_1+n_2)}{(n_1\mu_1+n_2\mu_2)^2+(\mu_1\mu_2B)^2(n_1+n_2)^2}, ~~~~\rm{and}~~~~
\end{equation}

\begin{equation}
R_{xx}(B)=R_s(0)\left[1+\frac{(n_1\mu_1n_2\mu_2(\mu_1-\mu_2)^2B^2}{(n_1\mu_1+n_2\mu_2)^2+((n_1+n_2)\mu_1\mu_2B)^2)}\right].
\end{equation}

\begin{table*}[!ht]
\centering
\begin{tabular*} {\textwidth}{c @{\extracolsep{\fill}} c c c c c c c}
\hline\hline
& $n_1$ & $\mu_1$  & $n_2$  & $\mu_2$ & $R_s$ (fit) & $R_s$ (exp.)\\ 
& (cm$^{-2}$) & (cm$^2$V$^{-1}$s$^{-1}$) & (cm$^{-2}$) & (cm$^2$V$^{-1}$s$^{-1}$) & ($\Omega/ \square$) & ($\Omega/ \square$)\\
\hline\hline
 Pre-illumination & 4.4 $\times 10^{13}$  & 34 & 8.0 $\times 10^{10}$ &845 &3997 & 4000\\ 
 Post-illumination &  4.7 $\times 10^{13}$  & 135 & 2.4 $\times 10^{12}$ &1000 &714 & 708\\  
\hline
\end{tabular*}
\caption{Values of the carrier density $n_{1,2}$, mobility $\mu_{1,2}$, and the zero-field sheet resistance $R_s$ (fit) estimated from the two-channel Drude model fit to the Hall resistance data for pre- and post-illumination. The experimental value of $R_s$ is calulated using the relation $R_s = \alpha R_{xx}$, where $\alpha (\sim 1.5)$ is a numerical ratio between $R_s$ and $R_{xx}$ determined by van der pauw measurements on different contact configurations.\cite{Guduru2013b, guduru2014}}
\label{table:1}
\end{table*}

Since the negative magnetoresistance can not be described using the simple Drude model, we discarded the $R_{xx}(B)$ from this analysis. We fit the Hall resistance data for pre- and post-illumination to the Eq. (2) in such a way that the fitting parameters $n_{1,2}$ and $\mu_{1,2}$ could satisfy Eq. (1) where $R_s(0)$ is experimental value of the sheet-resistance at zero-field. The best-fit  to the experimental data are shown as solid red-colour lines in the Fig.\ref{fig:MR}(c), and the yield values of carrier densities and mobilities are listed in Table~\ref{table:1}. As one can see, the fittings are quite fair, and justify the simultaneous existence of low-mobility and high-mobility subbands. For pre-illumination, the density of low-mobility carriers ($ 4.4~\times~10^{13}$~cm$^{-2}$) is almost three orders of magnitude larger than that of high-mobility ones ($8.0~\times~10^{10}$~cm$^{-2}$). Interestingly, a large increament in the density of high-mobility carriers is observed post illumination. 
Following the requirements for observing quantum oscillations ($\mu B \geq 1$), it is quite obvious that the high-mobility electrons are responsible for oscillations in $R_{xx}(B)$ above $\sim 5$ T  (Fig.~\ref{fig:MR}b).  
In order to get further insight on high-mobility carriers, we investigate the quantum oscillations in the following section.

\subsection{Quantum oscillations}

\begin{figure}
\centering
\label{fig:QOo}\includegraphics[width=14 cm]{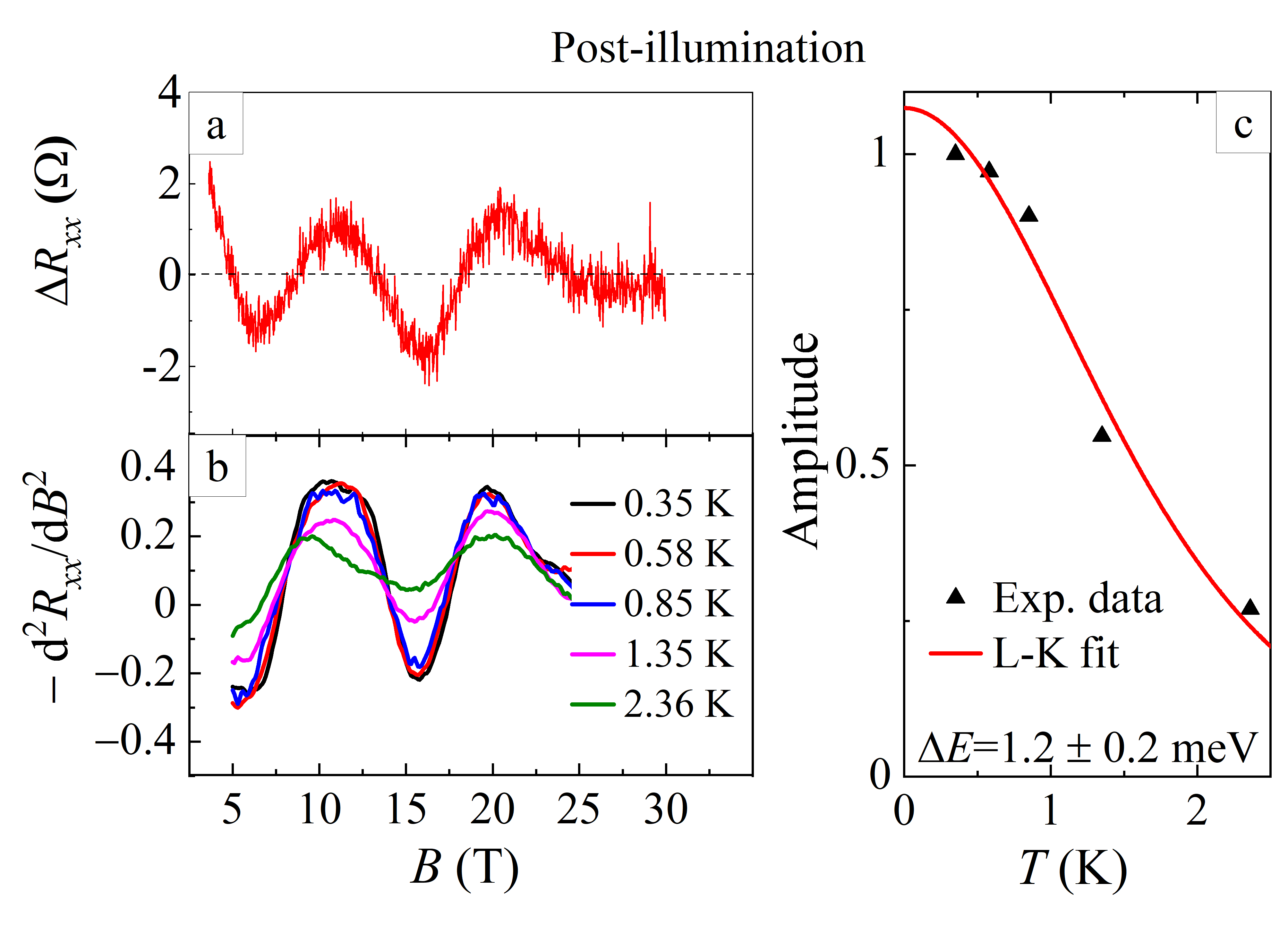}
\caption {(a) Magnetic field dependence of oscillating resistance ($\Delta R_{xx}$) after subtracting polynomial background from $R_{xx}(B)$ for pre- and post-illumination. (b) The second order derivative of the sheet resistance $R_{xx}$ (post-illumination) versus magnetic field  for five different temperatures.  (c) A normalized plot of the oscillation's amplitude (maximum-to-minimum difference for a oscillation's period in the field range of 13-25 T) with temperature. The fit is of the temperature dependent part of the Lifshitz-Kosevich formula and results in $\Delta E=1.2\pm 0.2$~meV.}
\label{fig:2ndDeriv}
\end{figure}

Despite comparable mobility of carriers for pre- and post-illumination, we perceived SdH oscillations only for post-illumination. We understood that the absence of oscillations for pre-illumination is because of small density of high-mobility carriers, which indicates a region of extreme quantum limit above $\sim$ 3 T. Fig. 3(a) shows the oscillating resistance ($\Delta R_{xx}$) after subtracting the polynomial background (second order polyomial fit in Fig. 2(b)) from the $R_{xx}(B)$ for post-illumination. The oscillations of amplitude $\sim 0.2-0.3~\%$ of $R_{xx}$ start developing at $\sim 5$ T field and vanish at $\sim$ 25 T. 
For cyclotron mass analysis, the post-illumination $R_{xx}(B)$ is measured at a few different temperatures and the second  derivative of $R_{xx}(B)$, $-$d$^2R_{xx}$/d$B^2$ is displayed in Fig.~\ref{fig:2ndDeriv}(b). To analyze the oscillations clearly, we have smoothed the $R_{xx}(B)$ data before taking the second derivative. We point out that the oscillations' frequency is similar in both cases, $\Delta R_{xx}(B)$ in Fig. 3(a) and $-$d$^2R_{xx}$/d$B^2$ in Fig. 3(b). The $R_{xx}(B)$ and $R_{yx}(B)$ are of the same order of magnitude, which means the matrix inversion relates a maximum in the resistance oscillations to a maximum of the conductivity, which is then proportional to the density of states (DOS). 

A straightforward interpretation of an oscillating DOS in a magnetic field is Landau quantization leading to quantum oscillations, such as SdH oscillations in the magnetoresistance. 
As high-mobility LAO/STO samples have been reported to show quantum oscillations \cite{McCollam2014,Fete2014a,rubi2020}, we have attempted to apply standard SdH oscillation analysis to our data. 
The temperature dependence of the oscillations' amplitude in Fig.~\ref{fig:2ndDeriv}(b) shows a decrease with increasing temperature following the expected Lifshitz-Kosevich behavior:\cite{Lifshitz1956}
\begin{equation}
\begin{split}
\label{eq:partLK}
\frac{\rho_{xx}}{\rho_0}(B,T)&= 1- 2e^{-\beta} \frac{\alpha T}{\sinh \alpha T} \cos \Big( 2\pi \frac{hn}{2eB}+\phi \Big) \\
\text{with }\alpha&=\frac{2\pi^2k_B}{\Delta E}, \beta=\frac{\hbar \pi}{\Delta E \tau}
\end{split}
\end{equation}
Most generally, $\Delta E$ is the energy between two extrema in the DOS (i.e. Landau level separation $\hbar e B/m_c$ with $m_c$ the cyclotron mass), $\tau$ the inelastic scattering time and $n$ the 2D electron concentration.
Using Eq.~(4) we can extract the parameter $\alpha$ from our experimental data and relate it to $\Delta E$. 
The fit to the temperature dependent oscillation's amplitude in Fig.~\ref{fig:2ndDeriv}(c) yields $\Delta E=1.2\pm 0.2$~meV. 
Relating this to a Landau level splitting we extract a cyclotron mass of $1.7 \pm 0.3~m_e$. Because there is only one complete oscillation visible in our data, the period is difficult to evaluate. However, a rough estimate yields $\Delta 1/B=0.04$~T$^{-1}$, which corresponds to a frequency of $25$~T, and carrier density of $n_{SdH} = 2ef/h \sim 1.2\times 10^{12}$~cm$^{-2}$. This value of the carrier density is comparable to that for high-mobility electrons estimated from the two-channel model fitting. 

\section{Discussion}

In order to get an insight on the origin of low- and high-mobility carriers in the illuminated LAO/STO system, we compare experimental results with electronic subband properties predicted theoretically. From the density functional theory (DFT) calculations, the conduction band of the 2DES at the LAO/STO interface is dominated by d$_{xy}$, d$_{xz}$, and d$_{yz}$ orbitals originated from the Ti:3d($t_{2g}$) orbital. 
While the d$_{xy}$ band is isotropic ($m^*_x = m^*_y = 0.7 m_e$), both d$_{xz}$ and d$_{yz}$ bands are quite anisotropic in $k_x$-$k_y$ space (for d$_{xz}$, $m^*_x = 0.7 m_e$ and $m^*_y$ = 7-9 $m_e$; for d$_{yz}$, $m^*_x$ = 7-9 $m_e$ and $m^*_y = 0.7 m_e$).\cite{popovic2008, delugas2011, rubi2020} Considering the geometric mean, the average effective mass of electrons residing in the d$_{xz/yz}$ bands is estimated as $\sim$ 1.3-1.8 $m_e$. Since the SdH oscillations probe the $k$-averaged cyclotron mass, a direct comparison of the experimentally estimated cyclotron mass  $\sim$ 1.7 $m_e$ with the ones calculated theoretically suggests that the high-mobility electrons occupy heavy d$_{xz/yz}$ subbands. Interestingly, the smaller density of the high-mobility electrons, as listed in Table 1, also agrees well with the layer-resolved density of states predicted by DFT calculations \cite{delugas2011, rubi2020}. Just above the Lifshitz point where the d$_{xz/yz}$ subbands begin to populate,  a majority of electrons located at the interface-adjacent STO layer occupy lower-lying d$_{xy}$ subbands and a minority of electrons populate the d$_{xz/yz}$ subbands extending farther from the interface \cite{Joshua2012, Ruhman2014, delugas2011, rubi2020, smink2017}. This remark may appear in disagreement with an ARPES study on a high-density sample ($n_{tot}$~=~6.5~$\times$~$10^{13}$ cm$^{-2}$) for which the Fermi sheet area of d$_{xz/yz}$ is comparable to that for d$_{xy}$ subbands. \cite{cancellieri2014} However, the lower density ($n_{tot}$ = 4.9 $\times 10^{13}$ cm$^{-2}$) of our sample would lead to a reduction in the Fermi wave vector \cite{cancellieri2014, smink2017}, and subsequently, to the lower carrier density in the d$_{xz/yz}$ than the d$_{xy}$ subbands.
In contrast to what would be expected from the comparison of effective masses, we find the mobility of the d$_{xy}$ electrons to be smaller than that of the d$_{xz/yz}$ electrons, in agreement with previous reports \cite{Joshua2012, Fete2014a,Yang2016,rubi2020}. The low-mobility of the $d_{xy}$ electrons is most likely due to increased scattering in the interface-adjacent STO layers, mediated by, e.g., a large strain\cite{cantoni2012}, mixed Ti valence states\cite{lee2016,rubi2020}, or a relatively large concentration of oxygen vacancies\cite{schutz2017, chikina2021}.

These strikingly observations strongly indicate that the illumination dopes a large amount of high-mobility electrons in the d$_{xz/yz}$ subbands, which leads to the elimination of the Kondo-like features and negative magnetoresistance. However, the low-mobility electrons having large carrier density in the studied sample belong to the d$_{xy}$ subbands and are responsible for the magnetism at the LAO/STO interface, in support of the prior conclusion made through spectroscopic studies \cite{lee2013}. It is worth mentioning here that the experimentally observed cyclotron mass could also be explained by electron-phonon renormalization for light subband using a mass enhancement factor of $\sim$ 2.5 \cite{cancellieri2016}.

\section{Conclusion}
To conclude, we have shown that an as-grown magnetic LAO/STO sample can exhibit higher mobility electrons by illuminating with a laser of higher energy than the bandgap of STO. The high-mobility electrons  ensue the quantum oscillations in high magnetic fields and at low temperatures. The carrier density and effective mass of the high-mobility electrons estimated from quantum oscillations evidenced that these electrons occupy heavy d$_{xz/yz}$ subbands. These findings are in line with the quantum transport studies in the high mobility LAO/STO samples acheived by optimising growth conditions \cite{BenShalom2010, Caviglia2010,Yang2016,rubi2020}. 
Furthermore, this work establishes an important bridge between the high-field magnetotransport studies on the low-mobility magnetic LAO/STO samples \cite{Guduru2013, Guduru2013b} and the high-mobility capped SrCuO$_2$/LAO/STO samples \cite{McCollam2014}. 

\section{acknowledgments}
We thank A. McCollam for discussions and a careful reading of the manuscript. 
We acknowledge the support of HFML-RU/FOM and LNCMI-CNRS, members of the European Magnetic Field Laboratory (EMFL). This work is part of the DESCO research program of the Netherlands Organization for Scientific Research (NWO). \\

\addcontentsline{toc}{section}{References}

\begin{thebibliography}{10}

\bibitem{Ohtomo2004}
A~Ohtomo and HY~Hwang.
\newblock A high-mobility electron gas at the {LaAlO$_3$/SrTiO$_3$}
  heterointerface.
\newblock {\em Nature}, 427(6973):423--426, 2004.

\bibitem{McCollam2014}
A~McCollam, Sander Wenderich, MK~Kruize, VK~Guduru, HJA Molegraaf, Mark
  Huijben, Gertjan Koster, David~HA Blank, G~Rijnders, Alexander Brinkman,
  et~al.
\newblock Quantum oscillations and subband properties of the two-dimensional
  electron gas at the {LaAlO$_3$/SrTiO$_3$} interface.
\newblock {\em APL Materials}, 2(2):022102, 2014.

\bibitem{Xie2014}
Yanwu Xie, Christopher Bell, Minu Kim, Hisashi Inoue, Yasuyuki Hikita, and
  Harold~Y Hwang.
\newblock Quantum longitudinal and hall transport at the {LaAlO$_3$/SrTiO$_3$}
  interface at low electron densities.
\newblock {\em Solid state communications}, 197:25--29, 2014.

\bibitem{Caviglia2008}
AD~Caviglia, Stefano Gariglio, Nicolas Reyren, Didier Jaccard, T~Schneider,
  M~Gabay, Stefan Thiel, German Hammerl, Jochen Mannhart, and J-M Triscone.
\newblock Electric field control of the {LaAlO$_3$/SrTiO$_3$} interface ground
  state.
\newblock {\em Nature}, 456(7222):624--627, 2008.

\bibitem{Gariglio2009}
Stefano Gariglio, Nicolas Reyren, AD~Caviglia, and Jean-Marc Triscone.
\newblock Superconductivity at the {LaAlO$_3$/SrTiO$_3$} interface.
\newblock {\em Journal of Physics: Condensed Matter}, 21(16):164213, 2009.

\bibitem{Richter2013}
Christoph Richter, Hans Boschker, W~Dietsche, E~Fillis-Tsirakis, Rainer Jany,
  Florian Loder, L~Fitting Kourkoutis, David~A Muller, John~R Kirtley,
  Christof~W Schneider, et~al.
\newblock Interface superconductor with gap behaviour like a high-temperature
  superconductor.
\newblock {\em Nature}, 502(7472):528--531, 2013.

\bibitem{Prawiroatmodjo2016}
Guenevere~EDK Prawiroatmodjo, Felix Trier, Dennis~Valbj{\o}rn Christensen,
  Yunzhong Chen, Nini Pryds, and Thomas~S Jespersen.
\newblock Evidence of weak superconductivity at the room-temperature grown
  {LaAlO$_3$/SrTiO$_3$} interface.
\newblock {\em Physical Review B}, 93(18):184504, 2016.

\bibitem{Brinkman2007}
Alexander Brinkman, Mark Huijben, M~Van~Zalk, J~Huijben, U~Zeitler, JC~Maan,
  Wilfred~Gerard van~der Wiel, GJHM Rijnders, David~HA Blank, and H~Hilgenkamp.
\newblock Magnetic effects at the interface between non-magnetic oxides.
\newblock {\em Nature materials}, 6(7):493--496, 2007.

\bibitem{Bert2011}
Julie~A Bert, Beena Kalisky, Christopher Bell, Minu Kim, Yasuyuki Hikita,
  Harold~Y Hwang, and Kathryn~A Moler.
\newblock Direct imaging of the coexistence of ferromagnetism and
  superconductivity at the {LaAlO$_3$/SrTiO$_3$} interface.
\newblock {\em Nature physics}, 7(10):767--771, 2011.

\bibitem{Gorkov2015}
LP~Gor'Kov.
\newblock Antiferromagnetism of two-dimensional electronic gas on
  light-irradiated {SrTiO$_3$} and at {LaAlO$_3$/SrTiO$_3$} interfaces.
\newblock {\em Journal of Physics: Condensed Matter}, 27(25):252001, 2015.

\bibitem{Bal2015}
VV~Bal, MM~Mehta, S~Ryu, H~Lee, CM~Folkman, CB~Eom, and Venkat Chandrasekhar.
\newblock Gate-tunable superconducting weak link behavior in top-gated
  {LaAlO$_3$-SrTiO$_3$}.
\newblock {\em Applied Physics Letters}, 106(21):212601, 2015.

\bibitem{Goswami2016}
Srijit Goswami, Emre Mulazimoglu, Ana~MRVL Monteiro, Roman W{\"o}lbing, Dieter
  Koelle, Reinhold Kleiner, Ya~M Blanter, Lieven~MK Vandersypen, and Andrea~D
  Caviglia.
\newblock Quantum interference in an interfacial superconductor.
\newblock {\em Nature nanotechnology}, 11(10):861--865, 2016.

\bibitem{Liang2013}
Haixing Liang, Long Cheng, Xiaofang Zhai, Nan Pan, Hongli Guo, Jin Zhao, Hui
  Zhang, Lin Li, Xiaoqiang Zhang, Xiaoping Wang, et~al.
\newblock Giant photovoltaic effects driven by residual polar field within
  unit-cell-scale {LaAlO$_3$} films on {SrTiO$_3$}.
\newblock {\em Scientific reports}, 3(1):1--7, 2013.

\bibitem{Xu2016}
Chencheng Xu, Christoph B{\"{a}}umer, Ronja~Anika Heinen, Susanne
  Hoffmann-Eifert, Felix Gunkel, and Regina Dittmann.
\newblock {Disentanglement of growth dynamic and thermodynamic effects in
  {LaAlO$_3$/SrTiO$_3$} heterostructures}.
\newblock {\em Scientific Reports}, 6:22410, mar 2016.

\bibitem{Caviglia2010}
AD~Caviglia, Stefano Gariglio, Claudia Cancellieri, Benjamin Sac{\'e}p{\'e},
  Alexandre Fete, Nicolas Reyren, Marc Gabay, AF~Morpurgo, and J-M Triscone.
\newblock Two-dimensional quantum oscillations of the conductance at
  {LaAlO$_3$/SrTiO$_3$} interfaces.
\newblock {\em Physical review letters}, 105(23):236802, 2010.

\bibitem{Fete2012}
Alexandre F{\^e}te, Stefano Gariglio, AD~Caviglia, J-M Triscone, and M~Gabay.
\newblock Rashba induced magnetoconductance oscillations in the
  {LaAlO$_3$-SrTiO$_3$} heterostructure.
\newblock {\em Physical Review B}, 86(20):201105, 2012.

\bibitem{Guduru2013}
VK~Guduru, A~Granados~del Aguila, Sander Wenderich, MK~Kruize, A~McCollam, PCM
  Christianen, U~Zeitler, Alexander Brinkman, G~Rijnders, H~Hilgenkamp, et~al.
\newblock Optically excited multi-band conduction in {LaAlO$_3$/SrTiO$_3$}
  heterostructures.
\newblock {\em Applied physics letters}, 102(5):051604, 2013.

\bibitem{DiGennaro2013}
Emiliano Di~Gennaro, Umberto~Scotti di~Uccio, Carmela Aruta, Claudia Cantoni,
  Alessandro Gadaleta, Andrew~R Lupini, Davide Maccariello, Daniele Marr{\'e},
  Ilaria Pallecchi, Domenico Paparo, et~al.
\newblock Persistent photoconductivity in 2d electron gases at different oxide
  interfaces.
\newblock {\em Advanced Optical Materials}, 1(11):834--843, 2013.

\bibitem{Lu2014}
Hong-Liang Lu, Liang Zhang, Xiu-Mei Ma, Gui-Jun Lian, Jin-Bo Yang, Da-Peng Yu,
  and Zhi-Min Liao.
\newblock Photoelectrical properties of insulating {LaAlO$_3$-SrTiO$_3$}
  interfaces.
\newblock {\em Nanoscale}, 6(2):736--740, 2014.

\bibitem{Liu2016b}
GZ~Liu, J~Qiu, YC~Jiang, R~Zhao, JL~Yao, M~Zhao, Y~Feng, and J~Gao.
\newblock Light induced suppression of kondo effect at amorphous
  {LaAlO$_3$/SrTiO$_3$} interface.
\newblock {\em Applied Physics Letters}, 109(3):031110, 2016.

\bibitem{wang2011}
X.~Wang, W.~M. L{\"{u}}, A.~Annadi, Z.~Q. Liu, K.~Gopinadhan, S.~Dhar, and
  T.~Venkatesan.
\newblock Magnetoresistance of two-dimensional and three-dimensional electron
  gas in {LaAlO$_3$/SrTiO$_3$} heterostructures: Influence of magnetic
  ordering, interface scattering, and dimensionality.
\newblock {\em Physical Review B}, 84(7):075312, aug 2011.

\bibitem{Guduru2013b}
V.~K. Guduru, A.~McCollam, A.~Jost, S.~Wenderich, H.~Hilgenkamp, J.~C. Maan,
  A.~Brinkman, and U.~Zeitler.
\newblock {Thermally excited multiband conduction in {LaAlO$_3$/SrTiO$_3$}
  heterostructures exhibiting magnetic scattering}.
\newblock {\em Physical Review B}, 88(24):241301(R), dec 2013.

\bibitem{Jost2015}
A~Jost, VK~Guduru, S~Wiedmann, JC~Maan, U~Zeitler, Sander Wenderich, Alexander
  Brinkman, and H~Hilgenkamp.
\newblock Transport and thermoelectric properties of the {LaAlO$_3$/SrTiO$_3$}
  interface.
\newblock {\em Physical Review B}, 91(4):045304, 2015.

\bibitem{Li2015c}
Y~Li, Y~Lei, BG~Shen, and JR~Sun.
\newblock Visible-light-accelerated oxygen vacancy migration in strontium
  titanate.
\newblock {\em Scientific reports}, 5(1):1--7, 2015.

\bibitem{Reyren2007a}
Nicolas Reyren, Stefan Thiel, AD~Caviglia, L~Fitting Kourkoutis, German
  Hammerl, Christoph Richter, Christof~W Schneider, Thilo Kopp, A-S
  R{\"u}etschi, Didier Jaccard, et~al.
\newblock Superconducting interfaces between insulating oxides.
\newblock {\em Science}, 317(5842):1196--1199, 2007.

\bibitem{sondheimer1947theory}
EH~Sondheimer and Alan~Herries Wilson.
\newblock The theory of the magneto-resistance effects in metals.
\newblock {\em Proceedings of the Royal Society of London. Series A.
  Mathematical and Physical Sciences}, 190(1023):435--455, 1947.

\bibitem{Hernandez2012}
T.~Hernandez, C.~W. Bark, D.~A. Felker, C.~B. Eom, and M.~S. Rzchowski.
\newblock {Localization of two-dimensional electron gas in
  {LaAlO$_3$/SrTiO$_3$} heterostructures}.
\newblock {\em Physical Review B}, 85(16):161407, apr 2012.

\bibitem{han2016}
K~Han, N~Palina, SW~Zeng, Z~Huang, CJ~Li, WX~Zhou, D-Y Wan, LC~Zhang, X~Chi,
  R~Guo, et~al.
\newblock Controlling kondo-like scattering at the {SrTiO$_3$}-based
  interfaces.
\newblock {\em Scientific reports}, 6(1):1--8, 2016.

\bibitem{guduru2014}
Veerendra~K Guduru.
\newblock {\em Surprising magnetotransport in oxide heterostructures}.
\newblock PhD thesis, Radboud University, 2014.

\bibitem{Fete2014a}
Alexandre F{\^e}te, Stefano Gariglio, Christophe Berthod, Danfeng Li, Daniela
  Stornaiuolo, Marc Gabay, and Jean-Marc Triscone.
\newblock Large modulation of the shubnikov--de haas oscillations by the {R}ashba
  interaction at the {LaAlO$_3$/SrTiO$_3$} interface.
\newblock {\em New Journal of Physics}, 16(11):112002, 2014.

\bibitem{rubi2020}
Km~Rubi, Julien Gosteau, Rapha{\"e}l Serra, Kun Han, Shengwei Zeng, Zhen Huang,
  Benedicte Warot-Fonrose, R{\'e}mi Arras, Etienne Snoeck, Michel Goiran,
  et~al.
\newblock Aperiodic quantum oscillations in the two-dimensional electron gas at
  the {LaAlO$_3$/SrTiO$_3$} interface.
\newblock {\em npj Quantum Materials}, 5(1):1--7, 2020.

\bibitem{Lifshitz1956}
I.M. Lifshitz and A.M. Kosevich.
\newblock {Theory of Magnetic Susceptibility in Metals at Low Temperature}.
\newblock {\em JETP}, 2(4):636, 1956.

\bibitem{popovic2008}
Zoran~S Popovi{\'c}, Sashi Satpathy, and Richard~M Martin.
\newblock Origin of the two-dimensional electron gas carrier density at the
  {LaAlO$_3$} on {SrTiO$_3$} interface.
\newblock {\em Physical review letters}, 101(25):256801, 2008.

\bibitem{delugas2011}
Pietro Delugas, Alessio Filippetti, Vincenzo Fiorentini, Daniel~I Bilc, Denis
  Fontaine, and Philippe Ghosez.
\newblock Spontaneous 2-dimensional carrier confinement at the n-type
  {SrTiO$_3$/LaAlO$_3$} interface.
\newblock {\em Physical review letters}, 106(16):166807, 2011.

\bibitem{Joshua2012}
Arjun Joshua, S~Pecker, J~Ruhman, E~Altman, and S~Ilani.
\newblock A universal critical density underlying the physics of electrons at
  the {LaAlO$_3$/SrTiO$_3$} interface.
\newblock {\em Nature communications}, 3(1):1--7, 2012.

\bibitem{Ruhman2014}
Jonathan Ruhman, Arjun Joshua, Shahal Ilani, and Ehud Altman.
\newblock Competition between kondo screening and magnetism at the
  {LaAlO$_3$/SrTiO$_3$} interface.
\newblock {\em Physical Review B}, 90(12):125123, 2014.

\bibitem{smink2017}
AEM Smink, JC~De~Boer, MP~Stehno, A~Brinkman, WG~Van Der~Wiel, and
  H~Hilgenkamp.
\newblock Gate-tunable band structure of the {LaAlO$_3$-SrTiO$_3$} interface.
\newblock {\em Physical review letters}, 118(10):106401, 2017.

\bibitem{cancellieri2014}
Claudia Cancellieri, Mathilde~L Reinle-Schmitt, Masaki Kobayashi, Vladimir~N
  Strocov, PR~Willmott, Denis Fontaine, Ph~Ghosez, Alessio Filippetti,
  P~Delugas, and Vincenzo Fiorentini.
\newblock Doping-dependent band structure of {LaAlO$_3$/SrTiO$_3$} interfaces
  by soft x-ray polarization-controlled resonant angle-resolved photoemission.
\newblock {\em Physical Review B}, 89(12):121412, 2014.

\bibitem{cantoni2012}
Claudia Cantoni, Jaume Gazquez, Fabio Miletto~Granozio, Mark~P Oxley, Maria
  Varela, Andrew~R Lupini, Stephen~J Pennycook, Carmela Aruta, Umberto~Scotti
  di~Uccio, Paolo Perna, et~al.
\newblock Electron transfer and ionic displacements at the origin of the 2d
  electron gas at the lao/sto interface: direct measurements with atomic-column
  spatial resolution.
\newblock {\em Advanced Materials}, 24(29):3952--3957, 2012.

\bibitem{lee2016}
PW~Lee, VN~Singh, GY~Guo, H-J Liu, J-C Lin, Y-H Chu, CH~Chen, and M-W Chu.
\newblock Hidden lattice instabilities as origin of the conductive interface
  between insulating {LaAlO$_3$} and {SrTiO$_3$}.
\newblock {\em Nature communications}, 7(1):1--8, 2016.

\bibitem{schutz2017}
P~Sch{\"u}tz, Dennis~Valbj{\o}rn Christensen, V~Borisov, F~Pfaff, P~Scheiderer,
  L~Dudy, M~Zapf, J~Gabel, YZ~Chen, Nini Pryds, et~al.
\newblock Microscopic origin of the mobility enhancement at a spinel/perovskite
  oxide heterointerface revealed by photoemission spectroscopy.
\newblock {\em Physical Review B}, 96(16):161409, 2017.

\bibitem{chikina2021}
Alla Chikina, Dennis~V Christensen, Vladislav Borisov, Marius-Adrian Husanu,
  Yunzhong Chen, Xiaoqiang Wang, Thorsten Schmitt, Milan Radovic, Naoto
  Nagaosa, Andrey~S Mishchenko, et~al.
\newblock Band-order anomaly at the $\gamma$-{Al$_2$O$_3$/SrTiO$_3$} interface
  drives the electron-mobility boost.
\newblock {\em ACS nano}, 15(3):4347--4356, 2021.

\bibitem{lee2013}
J-S Lee, YW~Xie, HK~Sato, C~Bell, Y~Hikita, HY~Hwang, and C-C Kao.
\newblock Titanium dxy ferromagnetism at the {LaAlO$_3$/SrTiO$_3$} interface.
\newblock {\em Nature materials}, 12(8):703--706, 2013.

\bibitem{cancellieri2016}
Claudia Cancellieri, Andrei~S Mishchenko, Ulrich Aschauer, Alessio Filippetti,
  Carina Faber, OS~Bari{\v{s}}i{\'c}, VA~Rogalev, Thorsten Schmitt, Naoto
  Nagaosa, and Vladimir~N Strocov.
\newblock Polaronic metal state at the {LaAlO$_3$/SrTiO$_3$} interface.
\newblock {\em Nature communications}, 7(1):1--8, 2016.

\bibitem{BenShalom2010}
M~Ben Shalom, A~Ron, A~Palevski, and Y~Dagan.
\newblock Shubnikov--de haas oscillations in {SrTiO$_3$/LaAlO$_3$} interface.
\newblock {\em Physical Review Letters}, 105(20):206401, 2010.

\bibitem{Yang2016}
Ming Yang, Kun Han, Olivier Torresin, Mathieu Pierre, Shengwei Zeng, Zhen
  Huang, TV~Venkatesan, Michel Goiran, JMD Coey, Ariando, et~al.
\newblock High field magneto-transport in two-dimensional electron gas
  {LaAlO$_3$/SrTiO$_3$}.
\newblock {\em Applied Physics Letters}, 109(12):122106, 2016.

\end{thebibliography}

\end{document}